# Highly Accurate, Reliable and Non-Contaminating Two-Dimensional Material Transfer System

Chandraman Patil[1], Hamed Dalir[1][2], Jin Ho Kang[3], Albert Davydov[4], Chee Wei Wong[3], Volker J. Sorger[1][2]*

[1]Department of Electrical and Computer Engineering, George Washington University, Washington, DC 20052, USA
[2]Optelligence LLC, 300 Delaware Ave Ste 210-A, Wilmington, DE 19801, USA
[3]Fang Lu Mesoscopic Optics and Quantum Electronics Laboratory, University of California, Los Angeles, CA 90095, USA
[4]Materials Science and Engineering Division National Institute of Standards and Technology Gaithersburg, MD, USA

*Corresponding Author: sorger@gwu.edu

## Abstract

The exotic properties of two-dimensional (2D) materials and 2D heterostructures, built by forming heterogeneous multi-layered stacks, have been widely explored across several subject matters following the goal to invent, design, and improve applications enabled by 2D materials. To successfully harvest these unique properties effectively and increase the yield of manufacturing 2D material-based devices for achieving reliable and repeatable results is the current challenge. The scientific community has introduced various experimental transfer systems explained in detail for exfoliated 2D materials, however, the field lacks statistical analysis and the capability of producing a transfer technique enabling; i) high transfer precision and yield, ii) cross-contamination free transfer, iii) multi-substrate transfer, and iv) rapid prototyping without wet chemistry. Here we introduce a novel 2D material deterministic transfer system and experimentally show its high accuracy, reliability, repeatability, and non-contaminating transfer features by demonstrating fabrication of 2D material-based optoelectronic devices featuring novel device physics and unique functionality. The system paves the way towards accelerated 2D material-based device manufacturing and characterization. Such rapid and material-near prototyping capability can accelerate not only layered material science in discovery but also engineering innovations.

# Introduction

Designing novel materials to improve the quality of life is appealing, but it is also challenging in practice. A generic solution for combining the best of different elements in one ultimate substance is still intricate and unsolved. Composite materials and III-V heterostructures have revolutionized many aspects of modern technologies laying the initial milestone in the process of engineering the material properties. However, it is still challenging and complex to mix and match crystalline materials with their unique properties of forming heterostructures with controlled tunning of attributes, functionalities, and properties. Developing novel techniques for using such materials like two (2D) dimensional materials is crucial for further material characterization enabling novel technological advancements.

The mechanical exfoliation of atomically thin 2D graphene layers from its bulk crystals in 2004 [1] introduced a unique research study of two-dimensional materials. This influenced the scientific community to investigate the fundamental graphene properties and applications in various domains [2], [3]. The invention of the micromechanical exfoliation method was the key to the success of graphene science, providing extremely high-quality samples and arousing curiosity for studying other 2D materials like transition metal dichalcogenides (TMDCs), MXenes, and 2D layered material-based insulators [4]-[11]. The exfoliation process produces crystal flakes of various sizes and thicknesses that are non-uniformly (i.e., spatially and material quality) and densely distributed over the target substrate leading to a low yield of useful atomically thin flakes. The hence required implementation of the optical identification method facilitates identifying useful atomically thin flakes on the substrate based on contrast difference in the image for different thicknesses allowing accurate and non-contact techniques of locating the flakes [12]. This method, however, is not suitable for fabricating complex systems such as heterogeneous integration of 2D materials on other pre-fabricated devices and structures [8] or building van der Waals (vdW) heterostructures by artificially stacking different 2D material flakes [13]-[18]. Therefore, it is important to develop new experimental setups and techniques for using 2D materials to fabricate more complex systems with enhanced reliability and functionality. Recently unique and modified techniques are developed such as the pick-and-place approach [19], the wedging method [20], the polyvinyl alcohol (PVA) method [21], patterned stamping method [22], viscoelastic stamping [23], and Evalcite method [24] for deterministic transfer of 2D materials. Each of these helps transfer 2D materials efficiently but is limited by their shortcomings and scope of implementation, such as obtaining a high yield of quality flakes, avoiding contamination of entire substrate from undesired flake transfer, demonstrating repeatability in device performance, and avoiding undesired residues on flakes [13].

The intrinsic properties of 2D materials are extremely sensitive to the chosen transfer and/or handling method deployed, which can involve temperature-assisted transfer, sacrificial polymer layers, litho-assisted patterned stamps, and/or wet chemical processes for substrate etching in the transfer process. The target substrate surface structures, for example, can be reactive to the chemicals used, the capillary forces involved in the process may crack very thin layers or produce capillary action-assisted undesired strain due to bubbles or wrinkling formed under surface in suspended material. 2D materials are also sensitive to temperature changes leading to surface oxidation of materials in few cases. Thus, the dry transfer methods, such as discussed in [22], [23], [25], help in overcoming wet-chemistry related challenges. Despite these modifications in improving the transfer method and system, the critical issue of spatial cross-contamination of the unwanted 2D material crystal flakes on the transfer substrate and yield of good quality flakes are yet to be addressed. An attempt of solving these issues was discussed in [13] focusing on overcoming the cross-contamination challenge in a 2D material transfer system by a comparison-based study. However, with increased research in the field of 2D materials, a reliable, robust, and highly repeatable method and transfer system is important for accelerating the 2D manufacturing process at free space, integrated electronic [26], [27], and/or photonic chip-level, straintronics [28] and large area optics applications.

Here we introduce a 2D material transfer system (herein termed *2DMTS)* capable of; i) eliminating contamination on 2D materials-based flakes, devices, and target substrate, ii) increasing the yield and quality of transferred material flakes, iii) enabling rapid and accurate transfer, and iv) achieving without involving wet chemical etching or thermal assisted process step for transfers. 2DMTS is a one-of-a-kind transfer method [13] using a viscoelastic polymer and metal micro-transfer stamper enabling selectivity during transfer. We demonstrate a reliable, robust, and highly accurate method to transfer 2D material onto arbitrary substrates with virtually no cross-contamination showing a high spatial accuracy of 0.5±0.2µm which is the highest reported to date without cross-contamination. After introducing 2DMTS and experimentally validating its capability and performance limitations, we put its rapid prototyping capability to use demonstrating novel device and fabrication capacities to include building and demonstrating optoelectronic devices featuring device novelty covering the fields of photonics, arrayed structural deposition of materials, heterostructure lattices, and electronic devices, including strainoptronics-based photodetector devices.

## Results and Discussion

The 2D transfer system platform (2DMTS) introduced and discussed here includes a metallurgical microscope, three linear axis stages for movement of Polydimethylsiloxane (PDMS) holder, a micro-transfer-stamper, and a target chip for precise alignment **(Fig. 1a)**. A small screw lock-based clamp needle holder is mounted on an angled arm XYZ axis stage to avoid blocking the field of view of a long working distance microscope mounted on top as seen in **Fig. 1a**. The micro-stamper can be lowered along Z-axis carefully to touch the substrate and transfer the selected flake after alignment as seen in **Fig. 1b**. The microscope is mounted with a standard 8-bit CMOS camera. It is used for the alignment of the system and measures the size and area of the flake before the transfer, calibrated for the specific objective lens. This allows for higher 2D material flake selectivity in terms of size and shape as per the requirement of the application. A camera software crosshair is used to align the position of flake, micro-stamper, and target using the respective stages. The soft PDMS helps in transferring material over different structures as it wraps around the profile providing better contact of material with the target without damaging it. The thin (17mil) PDMS sheets (Gel Pak) are cut down to small strips (1.5cm x 3cm) and are mounted on a microscope glass slide based PDMS holder (herein termed as PDMS cartridge). This cartridge is held by clamp mount metal clips keeping the PDMS suspended in the air, mounted on an XYZ axis stage as seen in schematic **Fig. 1c**. The clamp mount clips help to load and unload the

PDMS cartridge with ease (quick release) and prevent disturbing the alignment of other parts when changing the 2D material source banks. This allows multiple PDMS strips with different 2D material source banks to be realized for building heterostructures without the need of loading and unloading new PDMS films with different materials. The density of the flakes on Nitto tape is higher after exfoliation from a piece of crystal, forming clusters of un-isolated flakes. While transferring these flakes on a PDMS film, contacting the entire area of PDMS to the tape transfers clusters of 2D material flakes on the PDMS. The lower number of isolated flakes on PDMS leads to lower individual flake transfers from the 2D source bank cartridge. Therefore, it is important to transfer more isolated flakes on PDMS which are loosely adhered to the tape for ease of transfer and increasing the yield. A detailed description of the modified exfoliation of 2D material on PDMS film strips is discussed in the method below. The 2DMTS exfoliation graphical flowchart can be seen in **Fig. 1d** which shows that the high density of flakes and overlapping clusters formed on the exfoliation tape can be avoided or reduced significantly by using this modified method. The chip or target substrate is cleaned by rinsing with acetone and isopropyl alcohol (IPA) followed by nitrogen drying and is placed on an XY-θ stage where the rotation helps to load and unload the substrate for transfer of 2D material.

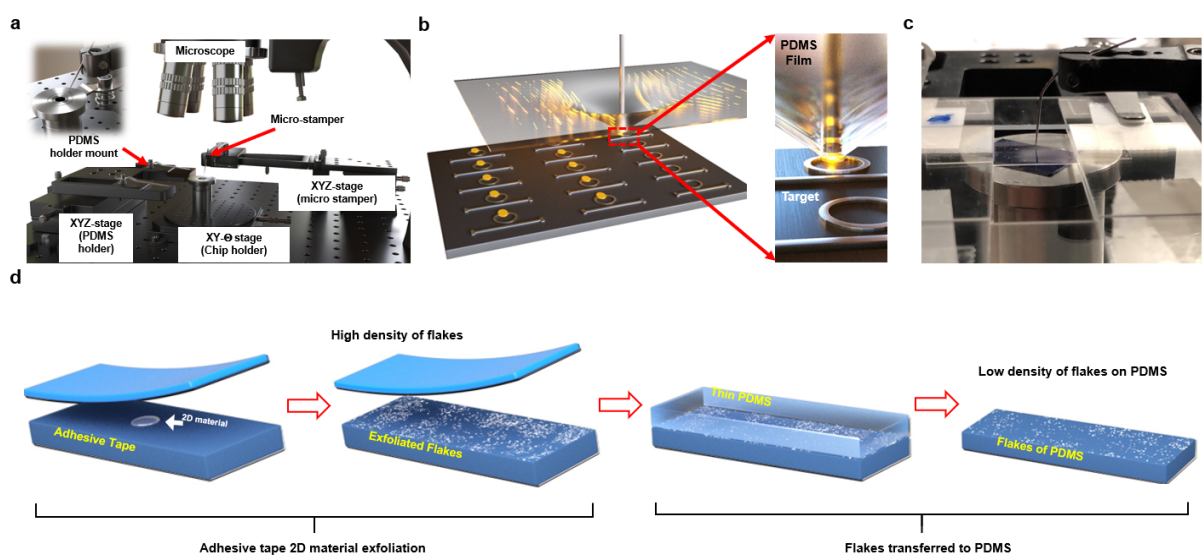

**Figure 1. Robust, precise, and reliable 2DMTS.** (a) The simplified 2D transfer system platform schematic includes the microscope, linear stages for alignment movement of PDMS holder, micro-stamper, and target substrate, PDMS holder mount clips and, a close view of the micro-stamper mounted on the setup. (b) Schematic of the PDMS, micro-stamper, and target chip position under alignment highlighting the close view of stretched PDMS when micro-stamper is approaching the targeted structure on the chip. Inset: showing the 2D material flake precisely positioned under the tip while the other flakes on the PDMS remain on the non-contact regions of the PDMS on the holder. The magnified view shows the small area of PDMS (largest reported here, ~400 μm x 200μm) under the stamper with the flake underneath touching the target. (c) Side angled image of the PDMS holder mounted on the stage with a PDMS strip mounted on it. The PDMS film is positioned over the target substrate and the micro-stamper is then adjusted on top of the selected flake as seen in the image. (d) The modified transfer process of 2D materials on PDMS using the adhesive tape exfoliation method helps to reduce the density of overlapping flakes and provides isolated flakes on PDMS for precisely locating flakes under the microscope. This allows selective transfer of the desired flake without cross-contamination on the chip.

The flake transfer yield is > (95.0±0.1) % **(Fig. 2)**, higher than any other state-of-the-art transfer system so far reported, as the entire surface area of the flake contacts with the target forming better adhesion to the target area. The failure in transferring the flake can be caused mainly because of the rough surface of the flake due to improper exfoliation, thus reducing the vdW adhesion strength with the target. A flake larger than the micro-stamper tip size reduces the transfer probability and is prone to break the flakes causing defects. But our different size micro-stamper range ensures flake-stamper size match. The contact area also defines the area of isolation (details in **Fig. 3** below) required for each flake on PDMS to avoid multiple flakes being transferred, here solved by using a modified exfoliation process as seen in **Fig. 1d** to reduce the density of flakes on the PDMS film and discussed further.

The 2DMTS setup has a transfer accuracy of about 0.7 µm and a precision of <100nm determined by experimental statistical analysis of the system. This analysis is based on independent experimental transfers of 80 flakes from different 2D materials (graphene, $MoS_2$, hBN, and $WSe_2$) for each stamper size by two different users. Thus, the dependency of a user experience, as observed in other transfer methods, is reduced by analyzing data generated by these two users who helped in the independent transfer of materials for producing data. The flake position distribution after transfer from the crosshair for the three stamper sizes can be seen in **Fig. 2a-c** where the inset shows the respective micro-stamper side view schematic. The distances of flakes transferred from the target point were measured using the calibrated camera software. The long working distance objective lenses can be switched to different magnifications for; 10x coarse (target search), 20x fine (flake search), and 40x finest (flake transfer) alignment based on the target size. Also, it is necessary to consider the flake size being transferred, therefore, for a wide range (1 um – ~150 um) of flake sizes, largest limited by the exfoliation method, we have built different elliptical tip-sized micro-stampers herein termed as stamper#1 (425µmx200µm), stamper#2 (300µmx145µm), and stamper#3 (185µmx130µm) built as discussing further. Details about the micro stamper are shared below. The histogram for the independent transfer attempts for each micro-stamper size shows the small offset, indicated by the peak position involved in the transfer location for varying micro-stamper sizes (**Fig. 2d**). The associated standard deviation distribution shows the dispersion of the transfer locations for each stamper normalized to the offset **(Fig. 2e)**. **Fig. 2f** shows the accuracy and precision of each micro-stamper demonstrating the highest accuracy of 0.7µm and precision of <100nm. Such accuracy and precision are vital in integrating the 2D material with highly dense photonic or electronic devices on-chip and nano-sensing applications. The smaller micro-stamper sizes provide higher accuracy and precision due to the small area of contact and lower offset deviation from the target. The maximum size of the flake that can be transferred and the effective area of contact of the PDMS with the three micro-stamper sizes considered in this article can be seen in **Fig. 2g**. This helps the user to choose the appropriate stamper size for achieving optimized transfer accuracy for desired flake size and target. The inset images represent the effective area of contact of PDMS to the target substrate when each flake is transferred and defines the area of isolation of flakes as well. The image contrast of the PDMS stamping area under the microscope is enhanced by the false color (purple) area for better representation.

For determining the yield of the transfer process produced by this system we transfer flakes of different sizes at different random locations on a $SiO_2$ substrate **(Fig.2h)**. This experimentation is designed to test the reliability and repeatability of the system for multiple single flake transfers that can be achieved in a single attempt. The transfer of the 2D material flake to the substrate without cross-contamination and 2D material deformation cracking of the flake into pieces by any means was termed as a success. The probability distribution of the success and failure rate of the transfer system was calculated using this data set as seen in **Fig. 2h**. All the 40 independent transfers were made in sequence without making any changes to the PDMS film or setup alignment apart from searching the different flakes on PDMS and changing the target location. Therefore, the setup is proved to be highly repeatable, reliable, precise, and accurate for a wide range of applications that can be developed for science and technological advancements. Such a plethora of system properties has not been achieved under any other 2DMTS as discussed in [13], [19]-[23], [29].

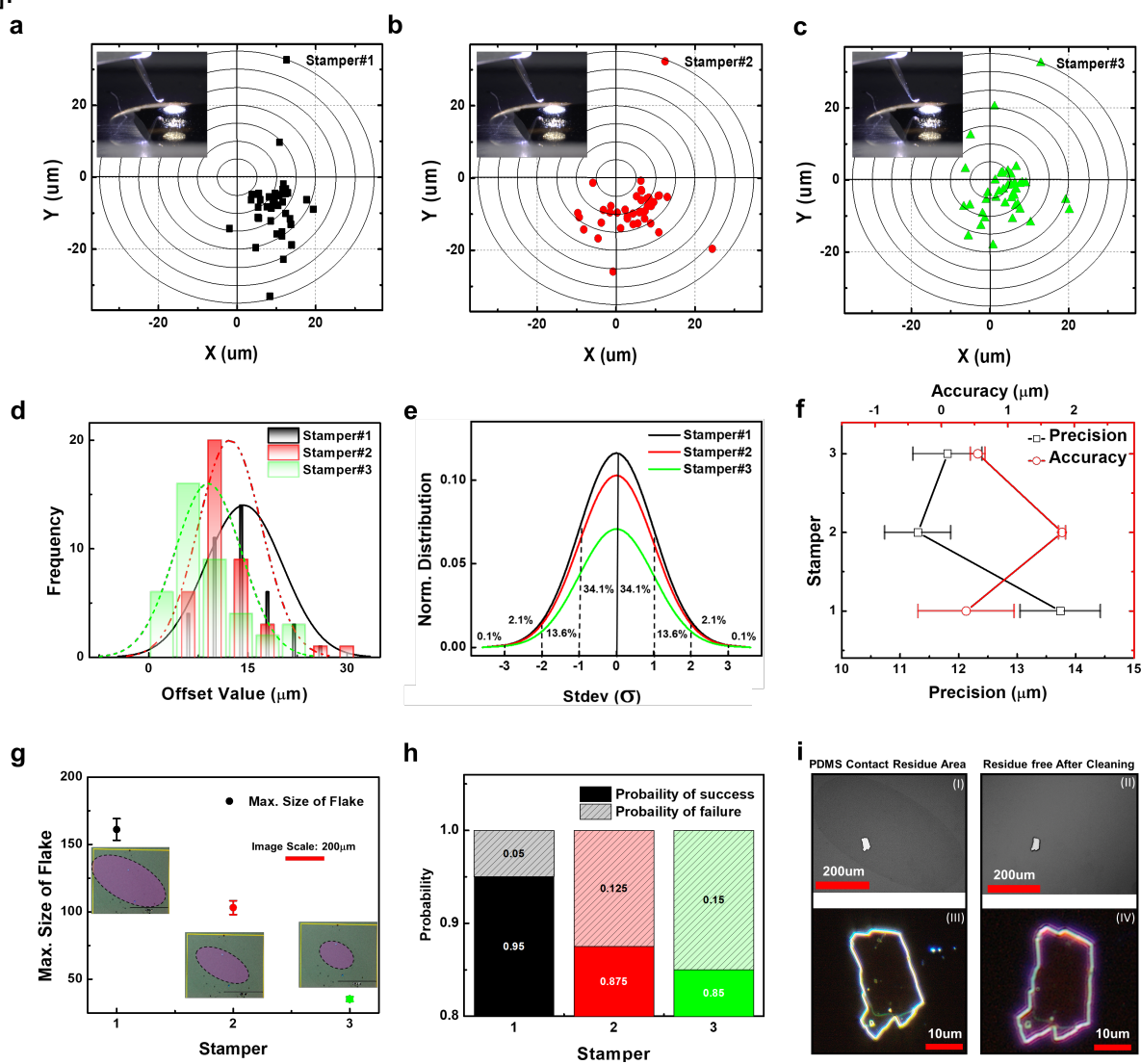

**Figure 2. 2D printer setup performance analysis for accuracy, precision, and reliability test.** (a)-(c) Distribution of position of flakes transferred on SiO$_2$ substrate measured from the center of the target. The offset produced due to optics and the size of the micro stamper can be observed in the graph. The inset shows the side view image representation of the micro stamper size and shapes positioned on top of the PDMS. (d) Histogram fitted with a gaussian curve that illustrates standard deviation of transferring process for each micro-stamper size. The graph also represents the offset in target location due to optics and stamper size (e) Standard deviation distribution of the flake transfer position on target measured considering the center of the flake to the software crosshair. Each band has 1 standard deviation, and the labels indicate the approximate proportion of area (note: these add up to 99.8% of data, and not 100 percent of data because of rounding for presentation). (f) The accuracy and precision values with error bars are computed for each stamper showing high accuracy for small stamper size and a reduced accuracy and precision for large stamper size. (g) Maximum size of flake that can be transferred with the three sizes of stamper. The inset microscope image represents the effective area of stamping for each stamper. The false-colored elliptical part highlights the stamping area where the scale bar is 200µm as the stamper shape is elongated in one direction as seen in Fig. 2 (a-c) inset (h) The probability distribution of transfer success and failure rate of the transfer system was calculated by gathering 40 data sets by transferring different flakes in a different location (i) Microscope image of transferred $MoS_2$ flake on oxide substrate with PDMS residue (I) and after cleaning process showing the residue-free area (II). The flake area after the transfer (III) and residue-free area after the cleaning process (IV) is confirmed by darkfield microscopy.

The flake after the transfer is cleared of PDMS residues on top of the flakes after the transfers have been completed. It is useful to consider samples where the material is transferred on a structured profile while rinsing and drying with nitrogen gas with as few vibrations as feasible. The vdW forces are prone to be not strong on the sides of the structures as the flakes may not conformally cover the structures owing to their thickness.

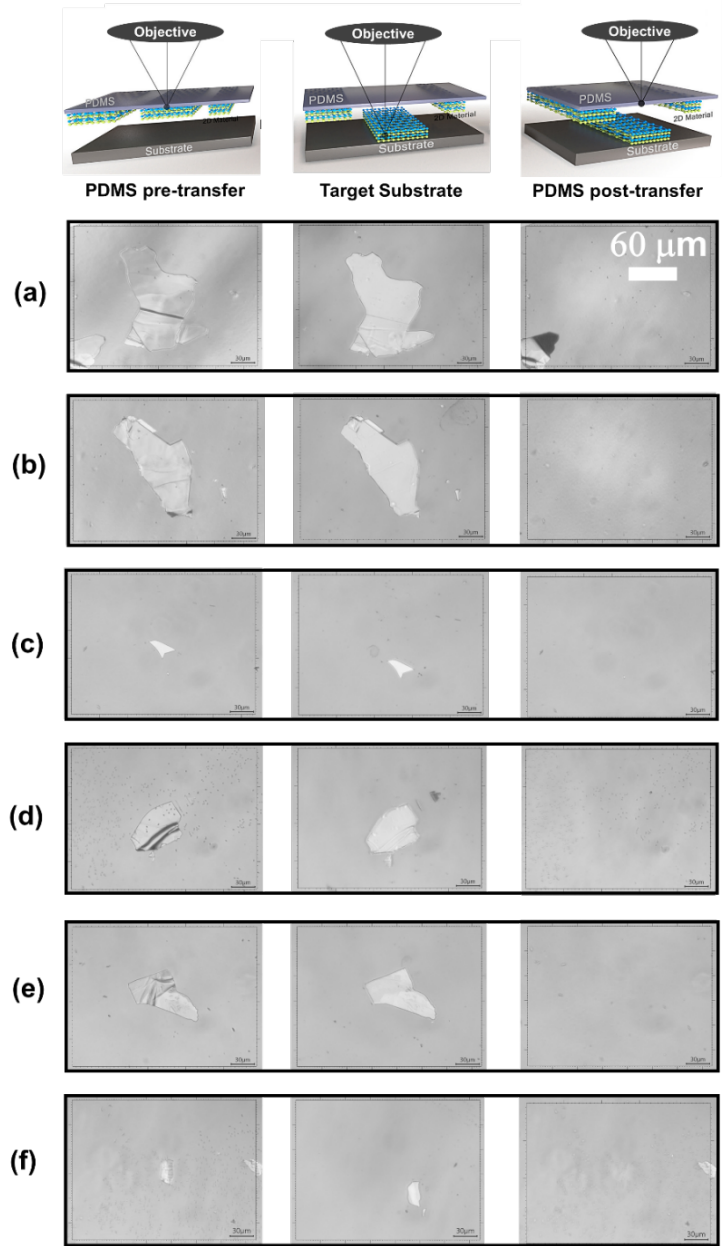

**Figure 3. Transfer results from the 2D material transfer system.** (a)-(f) Optical microscope images for 2D graphene flake on PDMS (before transfer), a substrate with flake after the transfer, and of PDMS (after transfer) captured while using the 2D printer setup. The images verify no cracking, wrinkles, or folding of the flake after transfer on the substrate.

This prevents the transferred flakes from moving off the surface due to mechanical vibrations. The issue of weak conformality of the flake coverage can be solved by keeping the sample in a vacuum chamber for 20-30 minutes. This strengthens the adhesion of the flake with the structure profile. On the sample, darkfield microscopy, optical filter-based microscopy, or secondary electron microscopy (SEM) can be used to check the results of the cleaning process depending on the resolution of inspection. As seen in **Fig. 2i**, a molybdenum disulfide ($MoS_2$) flake was transferred on $SiO_2$ substrate, and the contact area can be observed using a microscope, but the contrast is very poor for such small particles. Therefore, the area is enhanced by contrast correction which represents the area of contact in **Fig. 2i(I)**. The difference, in contrast, represents the contacted area of PDMS on the substrate and top of the flake. The residues are completely removed after the post-transfer simple cleaning process as seen in **Fig. 2i(II)**. The clean surface of the flake is verified using darkfield microscope images before **Fig. 2i(III)** and after the cleaning process **Fig. 2i(IV)**. This shows that the quality of the transferred flake is highly preserved along with no cross-contamination of flakes on the entire substrate.

      Further, understanding the non-cross contamination process while transferring multiple flakes in different locations on the substrate is necessary for repeatability and reliability demonstration. Also, there are possibilities of producing wrinkles or bubbles under the flake, cracking of large-sized flakes, and accidental transfer of a nearby flake after the transfer process. In **Fig.3a-f**, different sizes, and shapes of graphene material flakes are transferred on a $SiO_2$ substrate showing the optical images of flake on PDMS, flake transferred on the substrate, and of PDMS at the same location after transfer captured while using the 2D printer. The isolation of a nearby flake from the target flake while transfer can be observed in **Fig. 3a, f** where the flake on the side does not touch the substrate during the transfer process and can be seen left on the PDMS after transfer of the target flake. However, small flakes within the range of few microns cannot be avoided from getting transferred as it is difficult to isolate such small distances from cross-contamination as verified statistically in **Fig. 2g** to the stamper size. Further, these images show that the wrinkling or folding of material caused due to the viscoelastic nature of PDMS does not reflect after transfer on the substrate due to slight stretching on PDMS while stamping. This also indicates that the 2D material flakes are not strongly adhered to the PDMS due to the modified flake transfer technique discussed in **Fig. 1d** making it easy to transfer it on structures as well. Lastly, if small wrinkles are produced after the transfer, the sample can be kept in the high vacuum chamber to remove it.

## 2DMTS Preparation and Device Fabrication Process
### a) Micro-Stamper
      The micro-stamper is fabricated by modifying tungsten electrical probe needles. The needle is placed on the holder at 45 degrees, then lowered perpendicularly, and forced on the metal plate to bend the needle's tip. The amount of force given to the needle to bend the tip determines the size of the tip bent. The curved tip also provides a smooth tip characteristic to the micro stamper, preventing piercing in the PDMS sheet. The size of the stamping area

is calculated by measuring the PDMS contact area after transferring.

## b) Reusable PDMS Holder Preparation

The PDMS holder serves the purpose of holding multiple PDMS strips (here three) with a choice of having different/same material on each of the PDMS strips. The PDMS holder requires six thin glass slides (~1.2mm), Gel-Pak thin PDMS film (17 mils) strip/s, double-sided and single-sided adhesive tapes. The pictorial representation of preparing the PDMS holder is shown in **Fig 4. (1-5)**.

## c) Exfoliation Process

The exfoliation of 2D materials is a widely accepted technique using a weak adhesive Nitto tape discussed in [1]. To transfer the material onto the PDMS film, a pair of tweezers, weak adhesion Nitto exfoliation tape, a ruler, PDMS strips, and the desired 2D material crystals are needed. A small piece of crystal from bulk 2D material is placed on one end of a Nitto tape strip with a width equivalent to that of a PDMS strip. The other end is closed on the crystal and peeled back and forth until a lot of flakes spread on the tape.

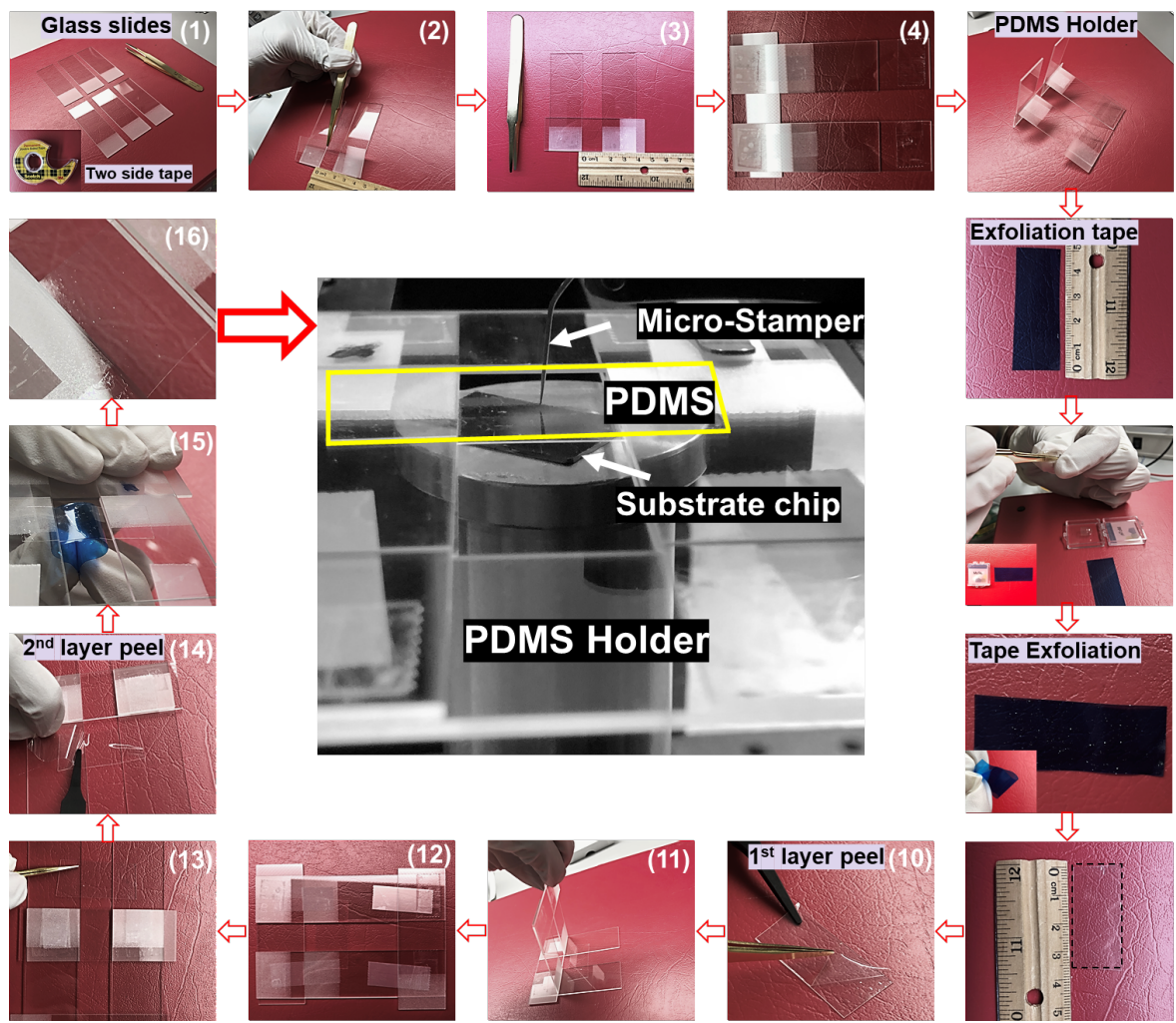

**Figure 4. Step-by-step pictorial flowchart for PDMS holder preparation and modified exfoliation technique.** The holder can be reused once prepared by replacing the PDMS film clamped in between the glass slide frame. The preparation of the holder requires thin glass slides (~1.2mm), Gel-Pak thin PDMS film

(17 mils), double-sided and single-sided adhesive tapes, pair of tweezers, weak adhesion exfoliation tape, ruler, and 2D material crystal. The figure in the center shows the placement of the holder on the stage for finding and positioning the 2D material flakes

The tape is usually then contacted with the substrate/target to transfer the flakes that have adhered to it. However, residues from the tape glue and other particles stuck to the tape are also added, resulting in a decreased yield of good-quality flakes. In addition, this spreads a lot of clusters of overlapping flakes all over the substrate, resulting in cross-contamination. The tape is gently pressed against the suspended PDMS film on the holder a few times. This permits the loosely bonded flakes on tape to adhere to the PDMS, reducing flakes clumping and allowing for more isolated flakes to be transferred. A complete pictorial procedure can be seen in **Fig. 4**.

### d) Transfer Process

The transfer process is simple and highly effective using the 2DMTS **(Fig. 5)**. The entire system is independent or requires minimal need for user experience and training. This makes this system more user-friendly as compared to other systems and avoids long training periods for a new user. The entire transfer process can be implemented in three simple steps: (i) Finding the flakes (ii) Positioning the flake and target area (iii) transferring the material. The desired size and shape of the flake can be selected on the PDMS by moving in the XY direction and using the measurement feature in the camera software. The microscope focal adjustment can be used to transfer the focus of vision to the micro-stamper, PDMS, and through the PDMS on target, once the PDMS holder is loaded on the stage. Using the micro-manipulating stages, a software crosshair assists in positioning the 2D material flake, target, and needle to a precise overlapping position. Following this alignment, the target and PDMS can be adjusted in the XY direction as needed to locate the flake and/or adjust the target's position. Once aligned to the center of the crosshair (homing position), the micro-stamper can be moved in the Z direction to make transfers. When PDMS touches the sample, the live video exhibits distinct indicators, allowing the operator to avoid over pressing the micro-stamper on the material while transferring. Once the PDMS has contacted the substrate, the operator can carefully bring the micro-stamper back to the homing position and repeat the operation for other flakes on the PDMS and target locations on the substrate. The micro stamper exerts stress of ~450kPa on the substrate and 2D flake while transferring simulated by ANSYS static structural simulation. This pressure is experimentally measured by placing a pressure sensor under the target substrate resulting in ~13kPa which is within the elastic limit of 2D material (GPa) [30], [31], and PDMS material (MPa) [32]. The simulation results for the stress and strain analysis of the 2DMTS on PDMS and substrate are discussed in detail in the supplementary section **(Fig. S1)**.

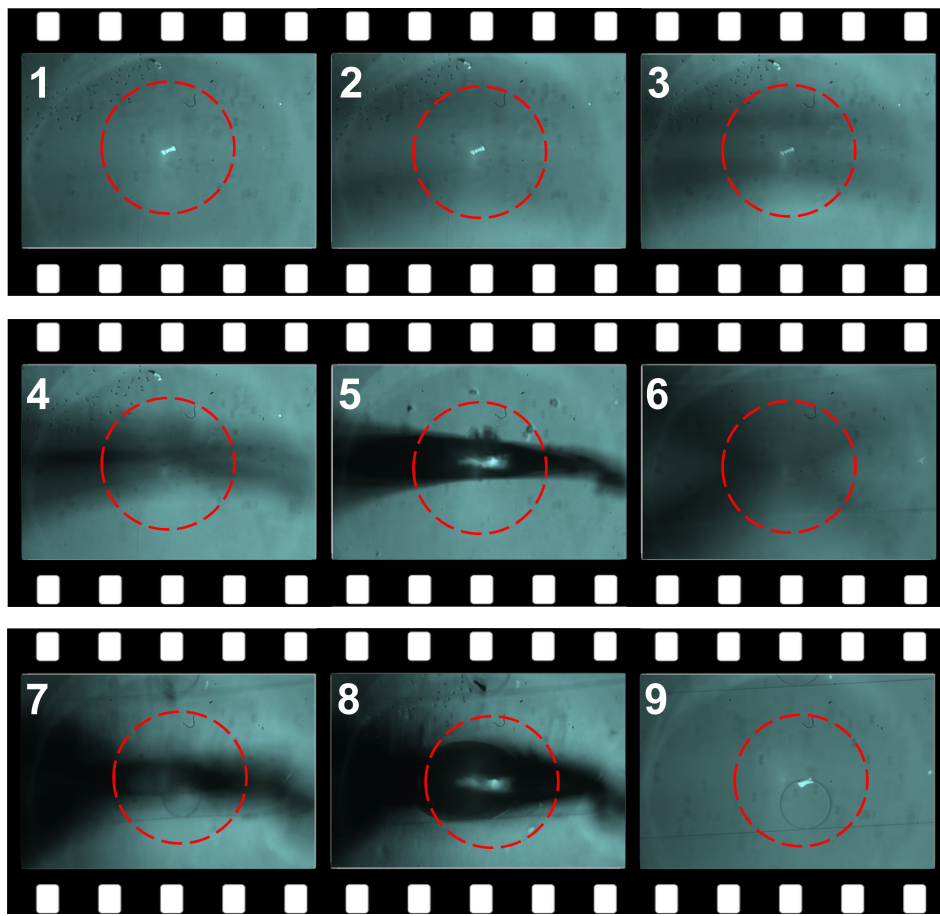

**Figure 5. The real-time video frames were acquired during the transfer of WSe$_2$ flakes onto a silicon Microring resonator structure.** The numbers on each frame denote the order in which the frames were captured during the transfer. Frames #1 - #4 show the micro stamper approaching the PDMS. Frame #5 shows the micro stamper tip contacting the PDMS, which continues to advance towards the target, while frames #6 - #9 show the stamper reverting to its original location.

**e) Nano-fabrication process**

The devices discussed in the application section in this article follow similar process steps unless otherwise specified further in the discussion. The study is performed on commercial Si photonic chips (Applied Nanotools INC.) tape-out. The flakes in all the applications discussed were transferred using the proposed system. The samples were rinsed clean using acetone followed by isopropyl alcohol for few minutes, both, before and after transfer. The electrical contacts were patterned using an e-beam lithography process assisted by pre-patterned alignment markers for positioning the flakes while making contacts. The Au/Ti (45nm/5nm) metal deposition was performed by e-beam evaporation technique followed by a lift-off process using acetone at room temperature.

**f) Electrical and optical measurement setup**

The experimental setup for measuring the heterogeneously integrated TMDC-Si devices comprises a tunable laser source (Agilent 8164B) and a broadband source (AEDFA-PA-30-B-FA) from where light is coupled into the grating coupler optimized for the propagation in the waveguide for 1550nm wavelength. The light output from the Si-MRR is coupled to the output fiber via a grating coupler and detected by a detector or an optical spectral analyzer (OSA202). A source meter (Keithly 2600B) was used for electrical measurements. A tunable (NKT SUPERCONTINUUM Compact) source and fixed wavelength laser diode module (CPS980 Thorlbas, INC.) at 980nm wavelength was used as sources for measuring the heterojunction based PN junction photodetector devices.

## Applications
**(a) Electrical Characterization of 2D materials post-2DMTS Transfer (Fig. 6)**

An extensive study has been performed on electrical properties of 2D materials, indicating the effect of impurities, trapped contaminants, and defects caused while transferring the material or post-process effects [33]-[35]. It is important to study the electrical properties of the material after contact formation for optimized performance as studied in [36]-[45]. However, achieving consistent results in the electrical properties of such devices is challenging for other transfer methods discussed earlier. To demonstrate consistency in the quality of flakes transferred using the proposed system we performed transmission line measurement on these flakes for sheet resistance measurement. The results were normalized by the area of the flake to remove the effect of the shape and size of flakes from the resistance measurement. Flakes with uniform thickness surfaces were considered for measurement confirmed by Raman spectroscopy before making the contact pads. As seen in the microscope image in **Fig. 6a** inset, $MoS_2$ flakes were transferred on $SiO_2$ substrate, and contacts were formed for different channels lengths from 0.5μm to 3.5μm with a step size of 0.5μm. This type of electrical characterization helps in designing the next-generation devices for improved optoelectronic device performances [46], [47]. **Fig. 6a** shows the variation in sheet resistance with varying channel length on the same flake representing the sheet resistance. Several such devices were fabricated at once by transferring multiple flakes on $SiO_2$ substrate for testing repeatability in achieving similar electrical properties. Repeatable electrical characterization was achieved by building 18 devices and observing their respective contact resistance ($R_C$ ~ <50kΩ) and sheet resistance ($R_S$ ~ <10kΩ/μm) of $MoS_2$ flake as seen in **Fig. 6b**. The electrical characterization of 2D materials is an important aspect to be studied for building electro-optic devices, magnetostatic devices, and stretchable electronic devices [47]-[50].

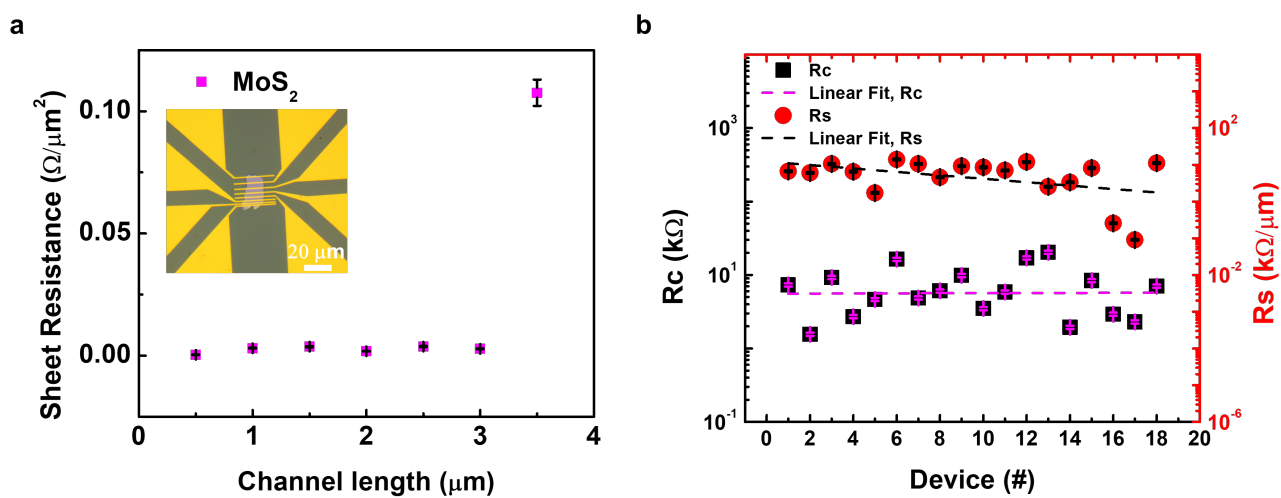

**Figure 6. Electronic resistivity test of 2D flakes transferred with 2DMTS.** (a) 2-terminal resistance measurement of individual $MoS_2$ flakes represented in the microscope image was performed to determine channel length-based sheet resistance of the material for electrical characterization. A microscope image of the device in the inset of $MoS_2$ flake transferred on $SiO_2$ substrate after designing the contacts for various channel lengths. (b) Comparative study for achieving repeatable electrical properties can be observed in the form of contact resistance ($R_C$ ~ <50kΩ) and sheet resistance ($R_S$ ~ <10kΩ/μm) for 18 devices.

**(b) Precise Transfer on an Array of Devices On-chip (Fig. 7)**

It is important to increase the yield of devices that can be fabricated for extensive research studies or commercial mass fabrication. A simple experiment is demonstrated here which proves the precise and accurate transfer capability of the system producing multiple devices on a chip in less than one minute spent on each device to transfer and few minutes in searching the desired flakes. The flake search time of few minutes can be minimized by prescanning the PDMS film and registering the flake positions using the top-mounted camera and a computer program by image processing. A commercial Si-photonic chip with arrays of micro-ring resonator (MRR) devices separated by 200µm distance was used in this experimentation. Each MRR was positioned under the system to transfer one flake in a sequence without changing any settings of the setup. The 2D material flakes, here $MoS_2$, were chosen at random and were targeted to be transferred at the same position on MRR for each device. Such transfers were made on 15 devices in a column on-chip of which three devices are shown in **Fig. 7a** with a close view for the transferred flake in **Fig. 7b**. Such robustness of the setup can be used to make devices for producing active and passive tunable devices [9], [51]-[53] with higher yield and repeatable results as discussed further.

**(c) Post passive tunning of Silicon**

The designed performance of silicon photonic circuits may vary due to fabrication uncertainties or incorporation of multiple post-fabrication process steps for fabricating active tunning devices on-chip. Also, tunning photonic circuits with precise control over the amount of tunning capability can be used for applications in neuromorphic computing [51], [52], [54]-[58]. A novel study was extensively discussed by [4] in past showing tunning of silicon microring resonator (MRR) coupling efficiency from under-coupled to the over-couple regime after the MRR was fabricated using $MoS_2$ using such stamping technique. Such studies can be crucial in understanding the effect of different 2D materials on optical properties after integration. Further helping to understand the optical properties of 2D materials like refractive index, optical absorption, phase change, and emission. We the proposed system such devices and study can be accelerated for studying any 2D material characterization study.

We here demonstrate a similar device as discussed in [4], showing enhanced coupling efficiency of Si-MRR from under coupled regime to critical coupling regime as seen in **Fig. 7c**. The MRR was transferred with two different small flakes as discussed in **Fig. 7** demonstrating precision in transferring flakes on devices extremely close to each other (here ~25um), as seen in **Fig. 7d**. The MRR spectral resonance enhancement was observed after transferring each of the two flakes to understand the change in coupling conditions by tunning the effective refractive index given by $(\Delta n)_{(eff,ring)} = ((2\pi R - L) * n_{(eff,bare)} + L * n_{eff}) / 2\pi R$ and seen in **Fig. 7e** [4]. Optical loss due to scattering from material edges and change in the imaginary part of the effective refractive index of the MRR was in total observed to be about 0.00251 dB/µm, represented in **Fig. 7f**.

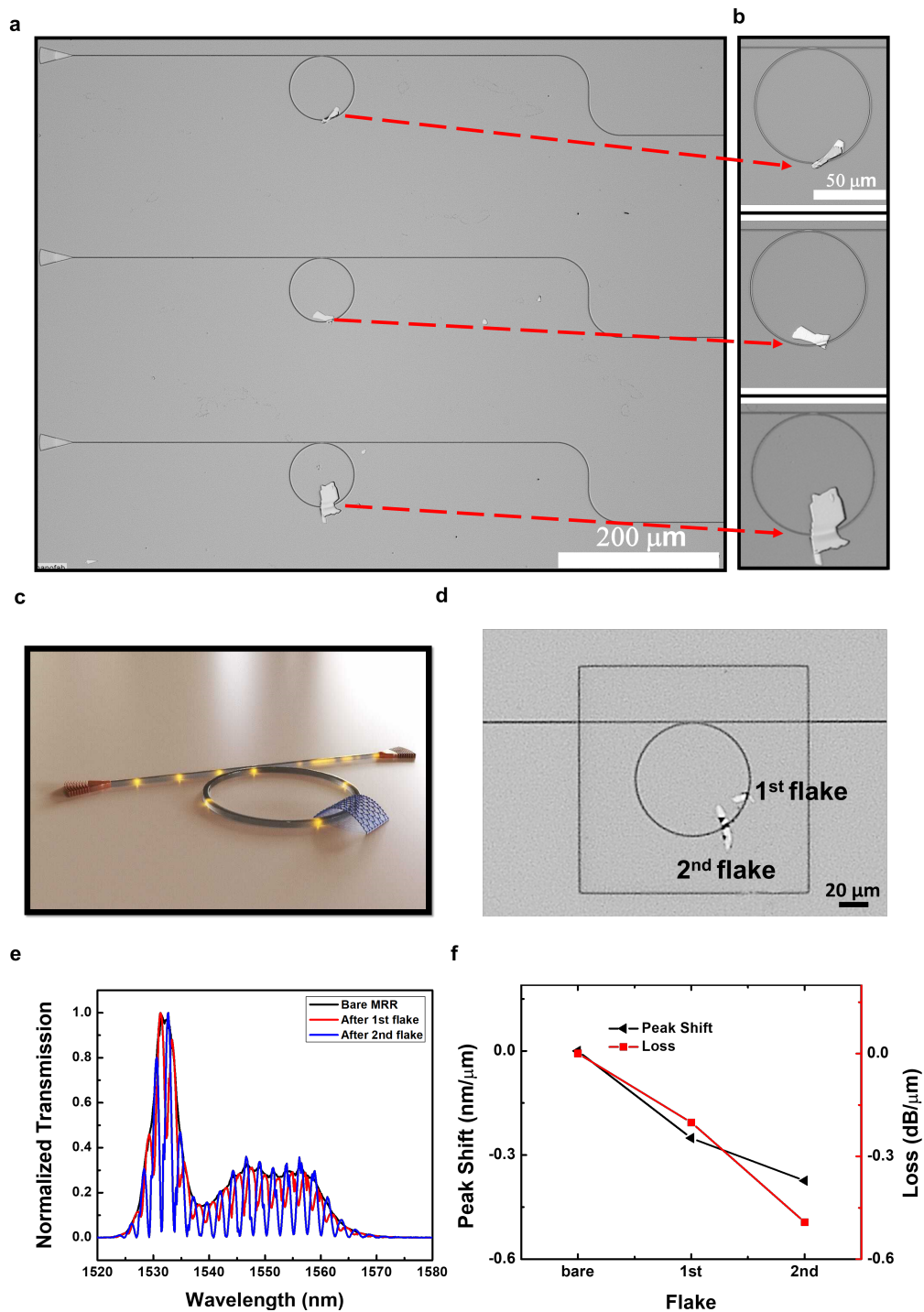

**Figure 7. 2DMTS performance test for accuracy, precision, and repeatability on assembling 2D structures atop photonic integrated circuit waveguides.** (a) Optical microscope images at 20x magnification of MoS₂ flakes transferred on Si-MRR devices in an array on-chip separated by 200nm distance from each other. (b) Zoomed in at 50x magnification to show the flake position on devices. **Silicon micro-ring resonator post coupling enhancement using MoS₂ thin flakes.** (c) Schematic (d) optical microscope image of an MRR (D = 80 µm & W = 500 nm) covered by two MoS₂ flakes precisely transferred using our developed 2D printer technique. (e) Spectral transmission output before and after the transfer of MoS₂ showing improvement of coupling efficiency. The improved resonance fringes can be observed in the graph (f) Resonance peak shift and optical loss after transfer of each flake can be observed showing linear absorption loss and variation in peak shift of resonance.

### (d) Strain Engineered MoTe$_2$ Integrated Photodetector on Silicon microring resonator (MRR)

We demonstrate a heterogeneous integration of 2D material using the proposed transfer system on a silicon photonic circuit on-chip using a few-layer MoTe$_2$ flake. Owing to the bandgap of MoTe$_2$, the optical absorption of the material is not suitable for 1550nm light detection. Using the proposed 2DMTS, the material can be strained on Si waveguides which enables bandgap tuning of MoTe$_2$. Therefore, allows the material to absorb at lowered bandgap due to induced strain. A single few-layer $MoTe_2$ flake was transferred on a silicon MRR based photodetector with 0.5A/W responsivity operating at 1550nm wavelength as discussed in [8]. A similar device has been designed and fabricated here for demonstrating the capability of the transfer system for realizing and studying such devices. The flake was transferred on a silicon MRR precisely using a micro-stamper and was patterned using the lithography process for contacts as seen in **Fig. 8a**. The microscope image as seen in **Fig. 8b** shows the device structure with electrical contacts. The current-voltage (I-V) characteristics of the device can be observed in **Fig. 8c**. A high responsivity of 0.02A/W (R = $I_{photon}$-$I_{dark}$ / $P_{in}$) was achieved for 2V electrical bias at 1550nm telecommunication wavelength. The dark current is observed around 50nA. A close view of the SEM image of the flake placed on the MRR can be seen in **Fig. 8d**. Multiple photodetector devices were studied here for different thicknesses, shapes, and sizes of the flakes, on the same chip with each MRR away from each other by 200 um showing the capability of array transfer represent in **Fig. 1b** to develop such devices with high repeatability. The responsivity of five such photodetectors at no bias (0V), 1V, and 2V can be seen in **Fig. 8e** integrated on Si-MRR with light coupled at 1550nm. The variation in performance for each device is due to variation in the coverage area of flake due to different sizes and variations in thickness. A detailed study of the effect of strain used to tune the absorption bandgap of $MoTe_2$ for such devices is discussed in [8] towards strainoptronics applications. These types of devices are promising for future on-chip detection and modulation of the signal without a complex fabrication process and high cost of production [6], [48], [65].

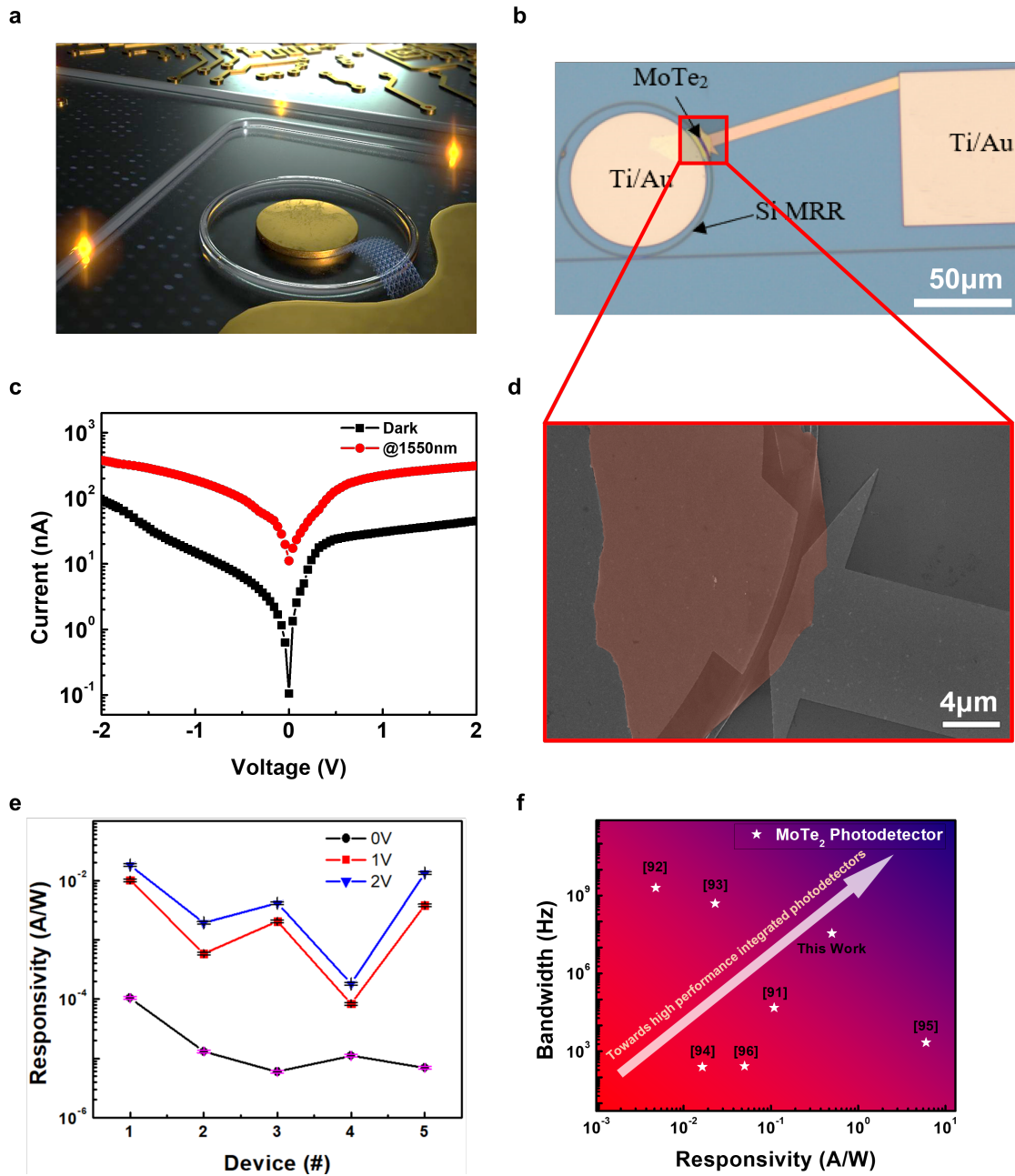

**Figure 8. A Si-MRR integrated strain engineered MoTe$_2$ photodetector.** a) Schematic representation of MoTe$_2$ based integrated photodetector for strainoptronics application built using the 2DMTS b) The structure of Si-MRR (radius (r)=40μm, height (h)=220nm, width (w)=500nm), and electrical contacts are represented in the optical microscope image. c) The I-V characteristics show the enhancement in current under the illumination of light coupled through Si-MRR at 1550nm wavelength. d) SEM image shows the coverage of flake on Si-MRR showing the coverage of the flake on the device e) Performance of five different photodetector devices represented by responsivity measured at zero bias (0V), 1V, and 2V showing enhancement in magnitude due to improvement in photogenerated carrier collection under influence of bias. f) Various photodetector devices discussed in [8], [59]-[64] are tabulated for understanding the bandwidth performance of $MoTe_2$ based photodetectors

**(e) Strain effect on $WSe_2$ bandgap stretched on nanostructures**

The atomically stacked 2D materials with each atomically thin layer adhered to each other under vDW forces show unique and strong tunning capabilities of modulating its properties by using mechanical forces. The properties of $WSe_2$ like photoluminescence(PL) enhancement, bandgap tuning, quantum emission, and optical absorption can be modulated by generating strain in the material [66]-[69]. Here we demonstrate the strain effect on $WSe_2$ multilayer using triangular geometric pillars. The pillars are etched from a SiN substrate of height 220nm. The structures are equally spaced to allow part of the flake to slack in between pillars and also stay suspended in few parts as seen in **Fig. 9a**. **Fig. 9b(I), (II)** show the microscope image of the nanopillars after the etching process and after a large $WSe_2$ flake transfer on the same area using the proposed transfer system respectively. In this configuration, it was possible to suspend the flake like a membrane on multiple nanopillars producing a high strain gradient along the pillar area. The corners of the triangular area help in strongly straining the membrane due to the virtue of the geometric shape. To understand the effect of induced strain, here, we performed photoluminescence and Raman spectroscopy spatial mapping. The Raman active modes of $WSe_2$ are represented in **Fig. 9c**. The PL spectra were analyzed by fitting the data to find the significant peak positions as seen in **Fig.9d**. The peak positions are highlighted in the figure at which the PL map analysis is performed for strain effects.

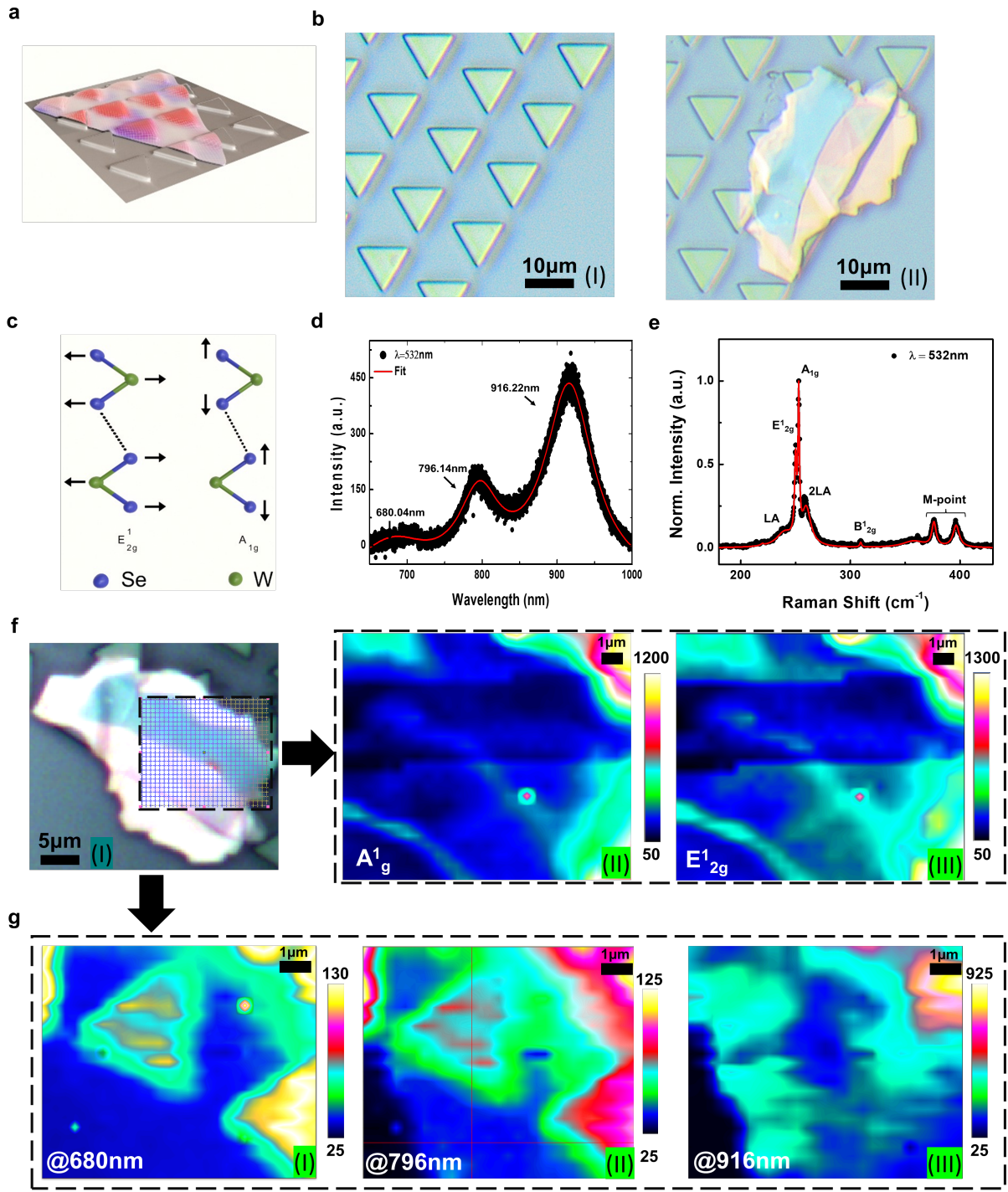

**Figure 9. Strain-induced in multilayer WSe$_2$ using triangular nanopillars.** (a) Schematic representation of strained multilayer WSe$_2$ membrane on triangular nanopillars. (b) Optical microscope images for (I) SiN triangular nanopillar structures (II) WSe$_2$ membrane (after transfer) using the 2D printer setup on the nanopillars. (c) Schematic representation of E$^1_{2g}$ and A$_{1g}$ Raman active modes of WSe$_2$. (d) PL spectra of WSe$_2$ membrane showing the fitted curve highlighting the significant active peaks. (e) Raman spectra showing the active modes ($\lambda_{excitation}$= 532nm). (f) Optical image of mapping area highlighted with the grid selection for Raman and PL mapping points (I), Raman intensity map for the strained area of the WSe$_2$ membrane at A$_{1g}$ mode (II), and E$^1_{2g}$ mode (III). (g) PL mapping realized at significant PL peaks for WSe$_2$ at 680nm, 796nm, and 916nm showing intensity enhancement for 680nm and 796nm peak position while a reduction in intensity at 916nm peak position.

The crystal lattice vibrations of bulk WSe$_2$ include Raman active modes of A$_{1g}$, E$_{1g}$, and E$_{2g}$ [70] **(Fig.9e)**. In Raman tensor analysis, the E-modes and A-modes are studied using circularly polarized helicity configuration [71]. In LR configuration (incident with left circular polarization and collect right circular polarization signal), we can get 247.4 cm-1 peaks consistently which denotes E$^1_{2g}$ mode peak, while in LL configuration (incident with left circular polarization and collect left circular polarization signal), we have 249.5 cm-1 peak which corresponds with A$_{1g}$ mode **(Fig. 9e)**. As seen in **Fig. 9f(I)**, an area was selected on top of few pillars with a small grid representing the mapping density using during the measurements. The effect of strain caused physical atomic arrangement variation in the crystal leading to change in vibrational modes of the material [66], [69]. **Fig. 9f** shows the Raman mapping profile of the WSe$_2$ membrane strained on a pillar showing the change of almost three times than unstrained region in the intensity of A$_{1g}$ mode (II) and E$^1_{2g}$ mode (III). The intensity variation here also shows the curved surface profile of the membrane on the pillars and the change in the peak intensity due to the strain introduced in the material. Further to test the optical response of the material under strained conditions, photoluminescence spectroscopy was performed to observe intensity enhancement at the significant peaks observed in **Fig. 9d**. For WSe$_2$, the PL spectra mapping was performed at 680nm, 796nm, and 916nm to understand the effect of strain on
each of them. Such studies can help in understanding the material properties to design and fabricate devices for exotic applications in various fields. The intensity map of PL for the strained region on one of the pillars at each of the significant peaks can be seen in **Fig. 9g**. An enhancement of about 4.5x and 3x was observed at 680nm (I) and 796nm (II) peak positions respectively. But a reduction in the intensity of about 2.7x is observed at 916nm (III). The selective tuning of the photoluminescence response of WSe$_2$ can is used for designing optical filters or spectral selective absorbers integrated on photonic circuits [72]-[74].

**(f) Building 2D material Heterostructures**

When we assemble numerous 2D crystals into a vertical stack, a multitude of possibilities emerge for exploring various properties exhibited by these materials. Such heterostructures, held together by vdW forces, allow significantly more combinations than any other standard material stacking growth process. The intricacy of the heterostructures that could be constructed with atomic accuracy is increasing as the family of 2D crystals grows [75]-[80]. Such 2D heterostructures can be used to build devices for electronic or electro-optic applications, superlattices, and quantum applications [50], [73], [77], [79], [81]-[96]. Here, we demonstrate a small example of building a MoS$_2$/WSe$_2$ heterojunction using the proposed 2D transfer system. A simple demonstration of combining individual unique properties of MoS$_2$ and WSe$_2$ by forming a junction from heterostructure. Both the MoS$_2$ and WSe$_2$ are transferred on a SiO$_2$ substrate to form the heterojunction as shown in Fig.10a. The inset in Fig. 10a shows the dark field microscope image of the heterostructure confirming the clean surface on the flakes and substrate. By experimental analysis of the unique material signature exhibited by Raman spectroscopy, we observe the combined effect of both the materials in a heterostructure. As seen in Fig. 10b, the Raman spectra for individual flakes of MoS$_2$ and WSe$_2$ can be observed with their active mode peaks. The MoS$_2$/WSe$_2$ heterojunction Raman spectra exhibit the Raman active modes of both the materials combined. The location for detecting the Raman signal is marked on the optical image in **Fig. 10a**. The data were corrected with background signal and baseline correction for removing the unwanted noise in the signal. This example shows the capability of the setup to realize heterostructure formation for different applications. The angle between each layered stack can be controlled in the setup for twistronics and superlattices applications as well but not discussed here.

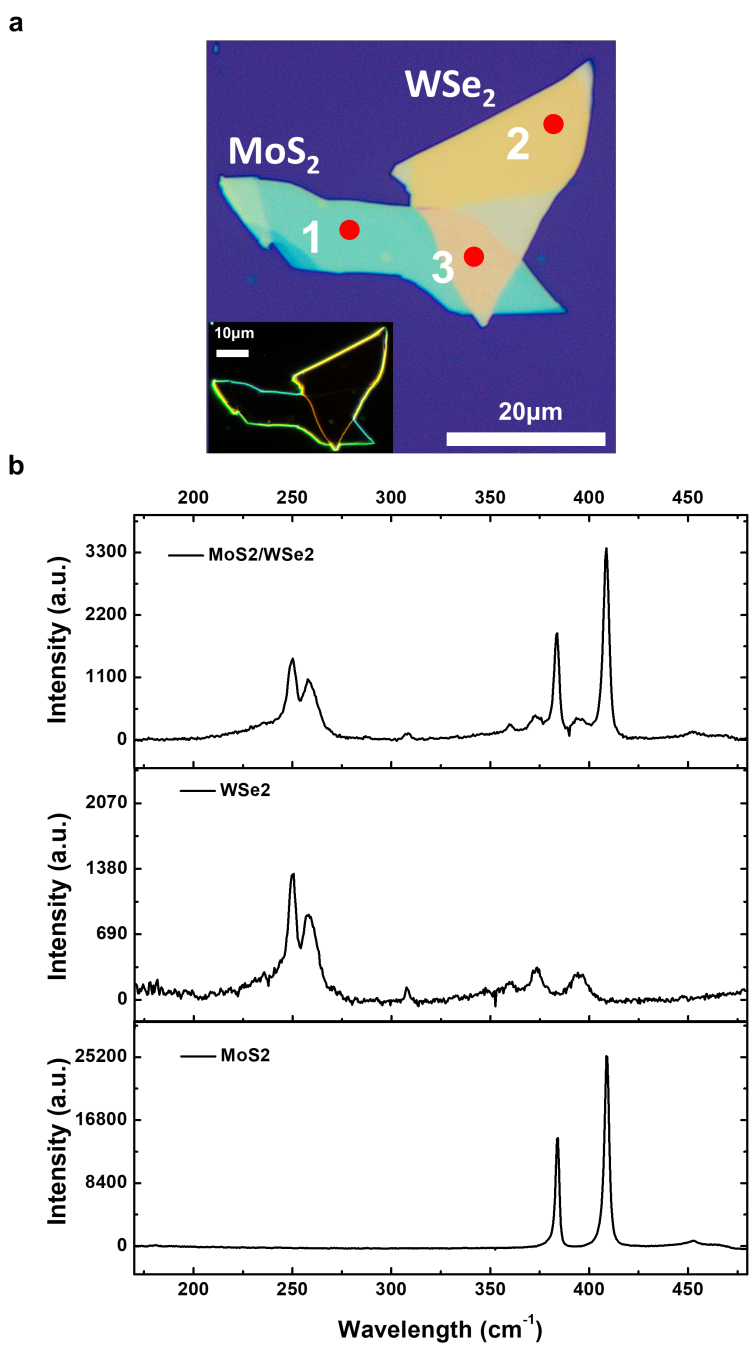

**Figure 10. MoS$_2$ and WSe$_2$ multilayer heterostructure.** (a) Optical microscope image of MoS$_2$/WSe$_2$ heterostructure built on SiO$_2$ substrate. The inset shows the dark field microscopy image of the heterostructure indicating no residues on the surface of the materials and substrate. (b) Raman spectra for MoS$_2$, WSe$_2$, and MoS$_2$/WSe$_2$ heterostructure showing the combined Raman signature of MoS$_2$ and WSe$_2$ at the junction formed by them.

## (g) InSe based PN-Junction Heterostructure Photodetector

Since past decades, 2D materials have been studied as promising photodetector materials, by changing the layer numbers or forming vdW heterostructures, owing to their high responsivity, fast response, broadband detection, low dark-current, and photo-detectivity [35], [65], [95], [97]-[107]. The operation of these high-performance devices demands high bias voltage leading to large power consumption. This limits technological applications in extreme environments, biomedical imaging, portable devices, etc. 2D indium selenide (InSe) has recently been investigated showing higher ultrasensitive photodetection characteristics [108] than other 2D semiconducting materials such as $MoS_2$ and $WSe_2$ [108]-[114]. Junction-based, e.g. p- and n- doped materials enabled realization in heterostructure devices by the formation of an atomically sharp p-n junction [94], [109], [115].

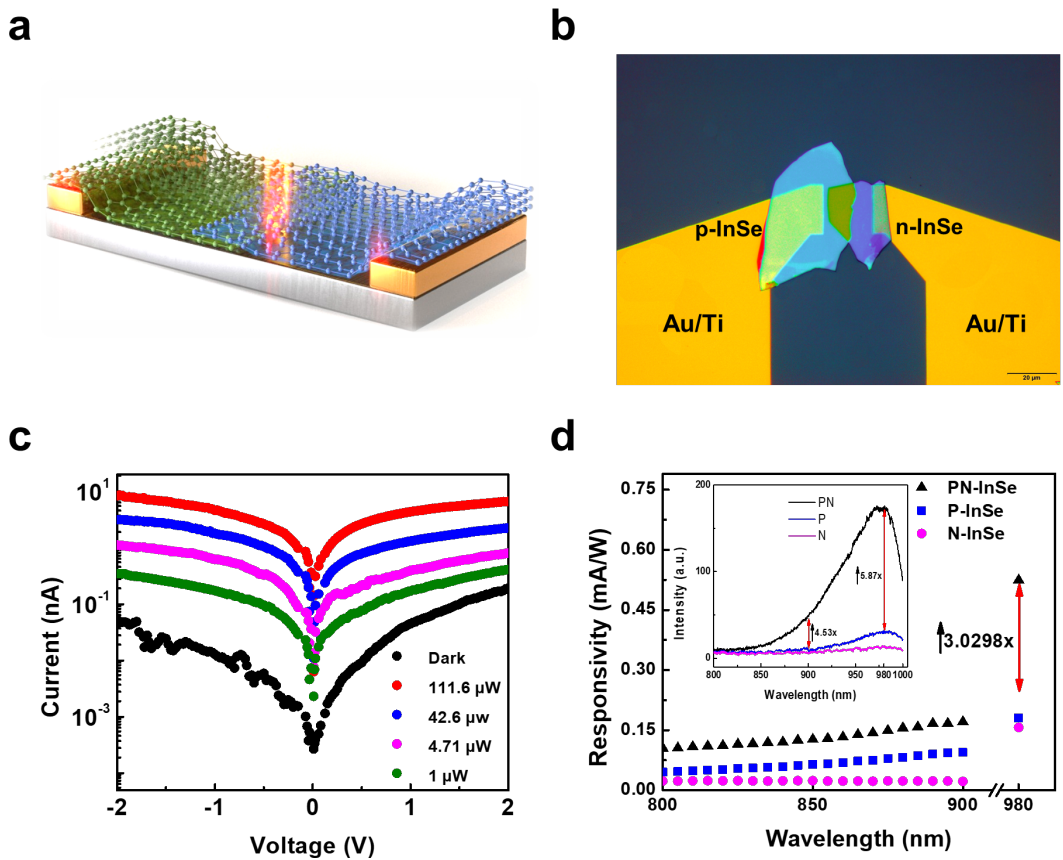

**Figure 11. 2D pn-junction heterostructure-based photodetector using 2DMTS.** (a) Schematic representation of the pn junction device. (b) Optical microscope image for 2D heterostructure stack of p- and n-doped InSe flake on $SiO_2$ substrate transferred on Au/Ti metal electrodes using the transfer system. (c) I-V characteristics of the device were tested at different optical power illumination at 980nm for photocurrent mapping. (d) The spectral responsivity of the pn junction device compared with control sample devices with n- and p-doped channel photodetectors showing enhancement of about 3.0298x. The inset shows the PL emission spectra intensity enhancement at the pn junction area compared to p and n regions.

Using 2DMTS, here we demonstrate a vDW heterostructure-based photodetector for near-infrared (NIR) spectrum absorption capable of efficient photo-detection operation at zero-bias enabled by a built-in voltage from the 2D material PN junction [116]. The important aspect of vDW 2D heterostructure is achieving a clean interface between the layers. In general, the quality of realizable heterojunctions is affected by chemical or mechanical degradation due to the presence of residues from adhesive from Nitto tape, polymer adhesives, or wet chemicals depending on the transfer method. By using a dry transfer medium and the mechanical stamping mechanism of 2DMTS this issue is eradicated. We heterogeneously integrated p-(Zn) and n-(Sn) doped InSe material using 2DMTS on pre-fabricated electrical contacts (Au/Ti) **(Fig. 11a)**. The sample was kept in the vacuum chamber for 30 minutes for better layer-layer and layer-metal adhesion. The optical microscope image seen in **Fig. 11b** shows the actual device configuration and structure of the electrical contacts. The center region forms the p-n junction formed by the heterostructure. The device's current-voltage characteristics were tested to observe dark current and photocurrent generation for 980nm wavelength illumination. The change in photocurrent for different optical power was also tested at 980nm showing a gradual increase in photocurrent as seen in **Fig.11c**. the spectral response of the pn junction photodetector was performed for a range of wavelengths from 800nm to 900nm using a tunable broadband source and also at 980nm using a laser module (VCSEL). A similar device with only n-doped and p-doped InSe channels was fabricated for comparison with the pn junction configuration of the device for the spectral response as seen in **Fig. 11d**. An enhancement of about 3x was observed in responsivity of about 0.524 mA/W for the pn-junction InSe photodetector as compared to only p- or n-doped InSe photodetector. The responsivity enhancement is due to the increase in the photoexcited electron-hole pairs between the p- and n- layer producing indirect excitons. These excitons have a higher carrier lifetime leading to enhanced performance of the device in photodetection [117]. The inset shows the PL emission for the p-, n-, and pn junction area of the device. An intensity enhancement of 5.9x and 4.5x is observed from the pn-junction at 980nm and 900nm wavelengths, respectively. This behavior aligns closely with the experimentally tested spectral response of the device. This signifies the generation of new strong absorption peaks in the heterostructurally built material lattice as compared to its natural material form. Such tunability of band absorption can be used for applications in the NIR spectra like LiDAR, gas sensing, photodetectors, optical modulators, and biosensing devices built using the proposed 2DMTS [11], [118]-[120].

**Conclusion**

In summary, the family of 2D crystals is continuously growing, both in terms of variety and number of materials as the scientific community progresses rapidly in enhancing the properties in these materials. In this work, we have successfully designed, developed, and demonstrated a novel robust transfer system for 2D materials for on-chip integration for building heterogeneously integrated devices and constructing heterostructures for applications like optoelectronic devices. Using a state-of-the-art micro-stamper and thin bendable film we were able to significantly improve the transfer of 2D materials reliably, with high repeatability, and without incurring any cross-contamination. The latter is a parasitic effect that is often caused by other transfer methods and eliminating this enables chip-industry like repeatability. We also demonstrate a diversity of applications for various active photonic, optoelectronic, and electrical devices, such as a zero-bias photodetector enabling pico-Watt level sensitivity whilst allowing for a dense integration on-chip with a repeatable transfer precision around 500nm, in the current implementation. This 2D material transfer system can be upgraded by automation, improving the accuracy, throughput, speed, and reproducibility. This robust and efficient transfer system provides a standard medium for expediting research and commercial large-scale integration of 2D materials towards making 2D materials-based integrated devices for a wide range of applications.

## SUPPLEMENTARY MATERIAL

Additional data analysis and simulation results are available in the supplementary document.

## CONFLICT OF INTEREST

The authors have no conflicts to disclose.

## ACKNOWLEDGMENT

VS is supported by AFOSR (FA9550-21-1-0095). A. V. D. acknowledges support through the Material Genome Initiative funding allocated to NIST.

## DATA AVAILABLILTY

The data that supports the findings of this study are available within the article [and its supplementary material].